\documentclass[aps,pra,noshowpacs,twocolumn,reprint,amsmath,amssymb,superscriptaddress,floatfix]{revtex4-1}

\usepackage{color}
\usepackage{amsmath}
\usepackage{graphicx}
\usepackage{dcolumn}
\usepackage{bm}

\begin{document}
\preprint{APS/123-QED}

\title{Direct Observation of Topology from Single-photon Dynamics on a Photonic Chip}

\author{Yao Wang}
\affiliation{State Key Laboratory of Advanced Optical Communication Systems and Networks, School of Physics and Astronomy, Shanghai Jiao Tong University, Shanghai 200240, China}
\affiliation{Synergetic Innovation Center of Quantum Information and Quantum Physics, University of Science and Technology of China, Hefei, Anhui 230026, China}

\author{Yong-Heng Lu}
\affiliation{State Key Laboratory of Advanced Optical Communication Systems and Networks, School of Physics and Astronomy, Shanghai Jiao Tong University, Shanghai 200240, China}
\affiliation{Synergetic Innovation Center of Quantum Information and Quantum Physics, University of Science and Technology of China, Hefei, Anhui 230026, China}

\author{Feng Mei}
\thanks{meifeng@sxu.edu.cn}
\affiliation{State Key Laboratory of Quantum Optics and Quantum Optics Devices, Institute of Laser Spectroscopy, Shanxi University, Taiyuan, Shanxi 030006, China}
\affiliation{Collaborative Innovation Center of Extreme Optics, Shanxi University, Taiyuan, Shanxi 030006, China}

\author{Jun Gao}
\affiliation{State Key Laboratory of Advanced Optical Communication Systems and Networks, School of Physics and Astronomy, Shanghai Jiao Tong University, Shanghai 200240, China}
\affiliation{Synergetic Innovation Center of Quantum Information and Quantum Physics, University of Science and Technology of China, Hefei, Anhui 230026, China}

\author{Zhan-Ming Li}
\affiliation{State Key Laboratory of Advanced Optical Communication Systems and Networks, School of Physics and Astronomy, Shanghai Jiao Tong University, Shanghai 200240, China}
\affiliation{Synergetic Innovation Center of Quantum Information and Quantum Physics, University of Science and Technology of China, Hefei, Anhui 230026, China}

\author{Hao Tang}
\affiliation{State Key Laboratory of Advanced Optical Communication Systems and Networks, School of Physics and Astronomy, Shanghai Jiao Tong University, Shanghai 200240, China}
\affiliation{Synergetic Innovation Center of Quantum Information and Quantum Physics, University of Science and Technology of China, Hefei, Anhui 230026, China}

\author{Shi-Liang Zhu}
\affiliation{National Laboratory of Solid State Microstructures and School of Physics,\\ Nanjing University, Nanjing 210093, China}
\affiliation{Synergetic Innovation Center of Quantum Information and Quantum Physics, University of Science and Technology of China, Hefei, Anhui 230026, China}

\author{Suotang Jia}
\affiliation{State Key Laboratory of Quantum Optics and Quantum Optics Devices, Institute of Laser Spectroscopy, Shanxi University, Taiyuan, Shanxi 030006, China}
\affiliation{Collaborative Innovation Center of Extreme Optics, Shanxi University, Taiyuan, Shanxi 030006, China}

\author{Xian-Min Jin}
\thanks{xianmin.jin@sjtu.edu.cn}
\affiliation{State Key Laboratory of Advanced Optical Communication Systems and Networks, School of Physics and Astronomy, Shanghai Jiao Tong University, Shanghai 200240, China}
\affiliation{Synergetic Innovation Center of Quantum Information and Quantum Physics, University of Science and Technology of China, Hefei, Anhui 230026, China}	
	
\date{\today}

\maketitle

\textbf{Topology manifesting in many branches of physics deepens our understanding on state of matters. Topological photonics has recently become a rapidly growing field since artificial photonic structures can be well designed and constructed to support topological states, especially a promising large-scale implementation of these states using photonic chips. Meanwhile, due to the inapplicability of Hall conductance to photons, it is still an elusive problem to directly measure the integer topological invariants and topological phase transitions for photons. Here, we present a direct observation of topological winding numbers by using bulk-state photon dynamics on a chip. Furthermore, we for the first time experimentally observe the topological phase transition points via single-photon dynamics. The integrated topological structures, direct measurement in the single-photon regime and strong robustness against disorder add the key elements into the toolbox of `quantum topological photonics' and may enable topologically protected quantum information processing in large scale.}\\

\noindent The introduction of topology into condensed-matter and material sciences originates from the connection of integer quantum Hall conductances with topological Chern invariants~\cite{topo_start}, which greatly expands our knowledge on state of matters. With the birth of topological insulators, searching topological state of matters in solid state materials~\cite{review_topo_1,review_topo_2} and photonic systems~\cite{Topo_review_1,Topo_review_2} has recently become a leading research field. In contrast to the challenging experimental requirements for realizing topological states in solid state materials, photonic systems provide a convenient and versatile platform to design various topological lattice models and study different topological states, including topological insulator states \cite{Hal2008,MITExp2009,Hafezi2011,PTI1,PTI2} and topological Weyl points~\cite{Topoto_weyl_1,Topoto_weyl_2}. The found topological boundary states potentially can be utilized for developing inherently robust and efficient artificial photonic devices~\cite{Topo_edge_1,Topo_edge_2,Topo_edge_3,QC1}.

In the view of fundamental physics, the topological invariant is a crucial parameter to characterize the topological matter state. In fermion systems, the topological invariant can be revealed by conductance measurements, while the concept of Hall conductance is inapplicable in photonic systems. New methods for directly detecting the topological invariants in topological photonics remain to be developed. The pioneering proposals in theory~\cite{BC2014,TTThe2009,CnThe2014} and experimental observations have been dedicated in both integrated photonic lattices~\cite{Berry2017,CnExp2015,TTExp2015} and bulk optics~\cite{exhqw,exnonhqw,exqwdisordernonh,exqwzak}.

Different from the efforts made to detect the topological invariant based on probing Berry curvature~\cite{BC2014,Berry2017}, non-Hermitian photon loss~\cite{TTThe2009,TTExp2015} or the dynamics of edge states~\cite{CnThe2014,CnExp2015}, we propose a new approach to directly detect the topological invariant via the bulk-state photon dynamics in the real space, which beyonds the physical picture where topological invariant is defined on the equilibrium Bloch state in the momentum space.

To extend promised topological protection into the quantum regime, we have to find an appropriate system that is physically scalable and has inherently low loss when scaling up. Integrated photonics can meet the first requirement elegantly by constructing topological structures on a photonic chip in a physically scalable fashion, with which topological states can be generated, manipulated and detected in a very high complexity beyond that conventional bulk optics can do~\cite{exhqw, exnonhqw, exqwdisordernonh, exqwzak}. Meanwhile, realizing topological states in Hermitian systems can well meet the second requirement since the intrinsic loss in non-Hermitian systems~\cite{TTExp2015, exnonhqw, exqwdisordernonh} 
will induce an evolution of exponential decay for single photons and multiplicative inefficiency for multi-photons.

Here, we integrate topological waveguide lattices on a photonic chip and experimentally demonstrate a direct observation of the topological invariants in the constructed Hermitian system using bulk-state photon dynamics. Through initially injecting photons into the middle waveguide to excite the bulk state, the values of topological winding numbers can be extracted from the chiral photonic density centers associated with the final output distribution. We further extend the topological system and measurement into quantum regime by observing the topological phase transition point via single-photon dynamics. With the bulk state excited by heralded single photons, we can successfully identify the topological phase transition point separating the topological trivial and nontrivial phases, even with artificially introduced disorder.\\

\begin{figure}[!t]
	\centering
	\includegraphics[width=1\columnwidth]{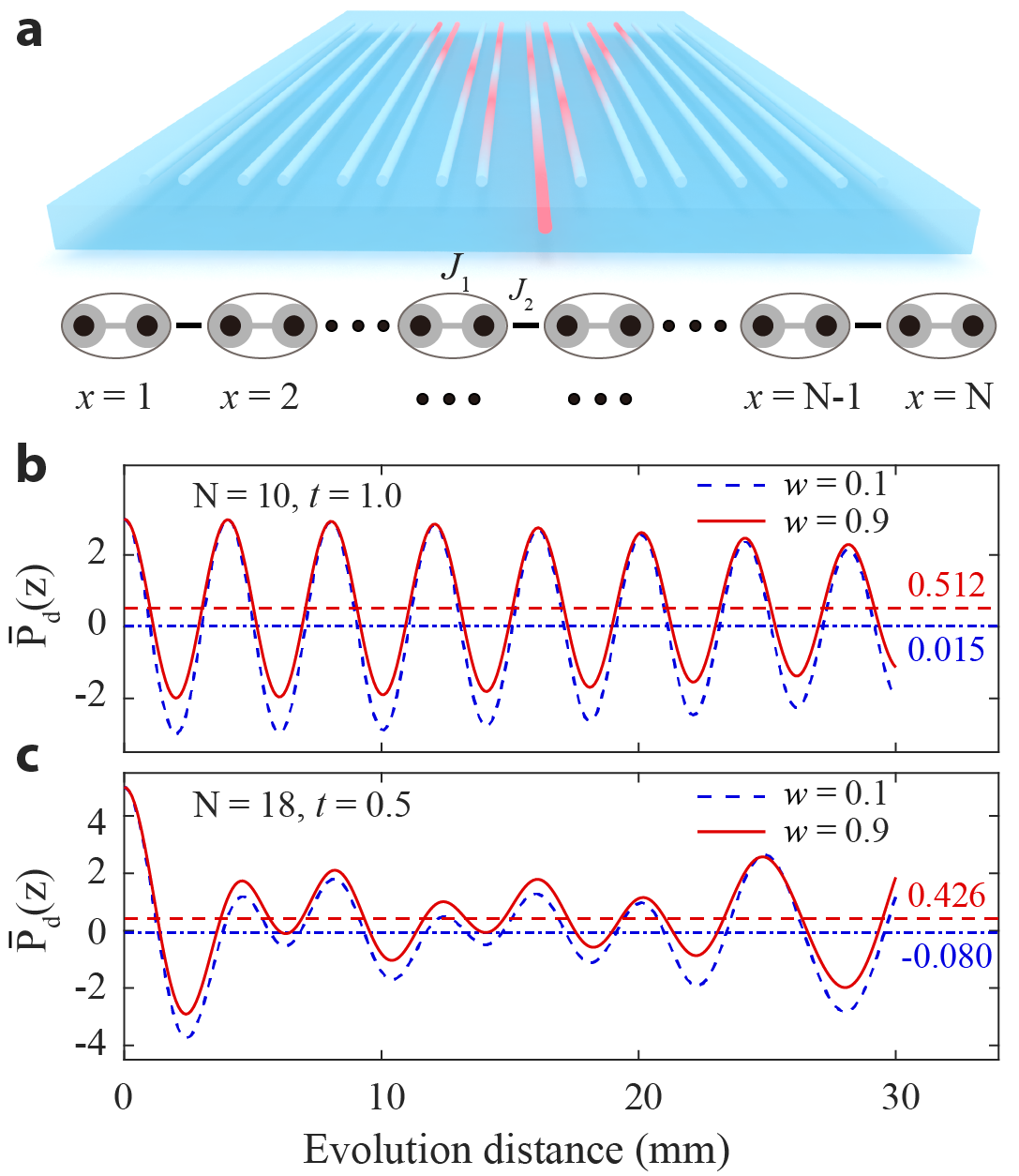}\\
	\caption{\textbf{Schematic of the integrated topological photonic lattice, model and simulation.} \textbf{a,} Sketch of Su-Schrieffer-Heeger model. The unit cell label $x$ is marked starting from the edge of the system with 1. Every unit cell consists of two sites and every site is implemented by a laser-written waveguide. \textbf{b-c,} Simulated results. The values of PPDC oscillate around 0 and 0.5 for $w=0.1$ and $w=0.9$ corresponding to the systems in topological trivial and nontrivial phase, respectively. The numbers on the right side of the figure present the  averaged values of PPDC.}
	\label{f1}
\end{figure}

\textbf{Topological photonic lattice}. As is shown in Fig.\ref{f1}(a), we fabricate waveguide lattices in borosilicate glass by using the femtosecond laser direct writing technique~\cite{fabri_1,fabri_2,fabri_3,fabri_4} (see Methods for details). The constructed lattices are based on the Su-Schrieffer-Heeger model, which describes a one-dimensional lattice with alternating strong and weak couplings. The Hamiltonian of this model could be written as~\cite{SSH}
\begin{equation}\label{SSH}
H=\sum_{x}(J_1a_x^+b_x+J_2b_x^+a_{x+1})+h.c.,
\end{equation}
where each unit cell in the chain consists of two sites labeled as $a$ and $b$, the terms $a^+_x$($a_x$) and $b^+_x$($b_x$) are the creation (annihilation) operators for the two sites in the $x$ unit cell, and the coefficients $J_1$ and $J_2$ represent the intra-cell and inter-cell coupling strengths, respectively. To study the topological feature, we rewrite Eq. (\ref{SSH}) in momentum space as
$\hat{H}=\sum_{k_{x}}\hat{h}(k_{x})$, where $\hat{h}(k_{x})=d_{x}\hat{\tau}_{x}+d_{y}\hat{\tau}_{y}$, $d_{x}=J_{1}+J_{2}\cos (k_{x})$, $d_{y}=J_{2}\sin (k_{x})$, and $\hat{\tau}_{x}$ and $\hat{\tau}_{y}$ are the Pauli spin operators defined in the momentum space. The energy bands of the Hamiltonian are characterized by the topological winding number
\begin{equation}
\nu =\frac{1}{2\pi }\int dk_{x}\mathbf{n}\times \partial _{k_{x}}\mathbf{n}
\label{wn}
\end{equation}%
where $\mathbf{n}=(n_{x},n_{y})=(d_{x},d_{y})/\sqrt{d_{x}^{2}+d_{y}^{2}}$. We manipulate the coupling coefficients as $J_1=g+gt\cos(w\pi)$ and $J_2=g-gt\cos(w\pi)$, where $g>0$, $0<w<1$, and $0\leq t\leq 1$. The system is in the topological nontrivial phase with the winding number $\nu=1$ when $J_1<J_2$, i.e. $w \in (0.5,1)$. Otherwise, it is in the topological trivial phase with $\nu=0$ when $J_1>J_2$, i.e. $w \in (0,0.5)$. The topological phase transition point appears when $J_1=J_2$.\\

\textbf{Dynamical detection of topological winding number}. To detect the winding number of the lattice on the topological photonic chip, we introduce a photon population difference center (PPDC) $P_d=\sum_{x}x(a_x^+a_x-b_x^+b_x)$, where the unit cell index $x$ is shown in Fig.\ref{f1}(a). We inject photons into the middle waveguide to excite the bulk state. With the evolution of the photons over a distance $z$ in the lattice, the corresponding PPDC can be denoted as $\bar{P}_d(z)$ (see Methods). We find that the topological winding number $\nu$ can be measured via the evolution-distance-averaged PPDC $\bar{P}_d(z)$, which can be expressed as (see Methods)
\begin{equation}\label{wnPPDC}
\nu=2\lim_{Z\to \infty}\frac{1}{Z}\int_{0}^{Z}dt\bar{P}_d(z),
\end{equation}
where $z$ is the evolution distance. In Fig.\ref{f1}(b-c), we calculate the $\bar{P}_d$ for different coupling coefficients and lattice sizes. The results show that the values of PPDC $\bar{P}_d$ keep oscillating centered at 0 and 0.5 when the lattice is in the topological trivial and nontrivial phases, respectively. The topological winding numbers derived as $\nu=0$ and $\nu=1$ can be directly measured from the output density distribution.

\begin{figure}[!t]
	\centering
	\includegraphics[width=0.96\columnwidth]{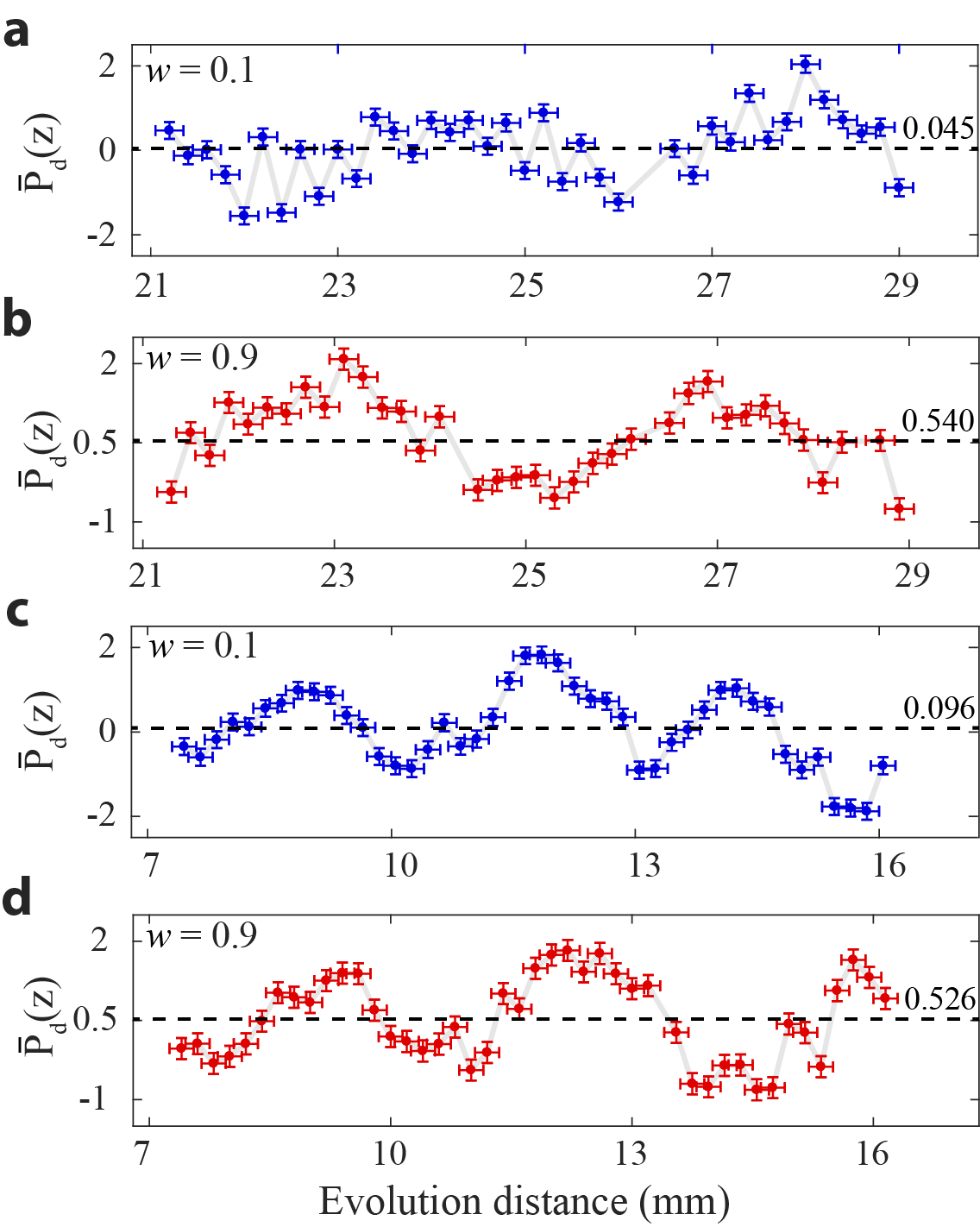}\\
	\caption{\textbf{Experimental results of PPDC.} The measured values of PPDC for 10-sited (\textbf{a, b}) and 18-sited (\textbf{c, d}) lattices, which are found oscillating around 0 (\textbf{a, c}) and 0.5 (\textbf{b, d}) for the systems in topological trivial and nontrivial phase. The averaged values of PPDC $\bar{P_d}$ are $0.045\pm 0.090$ (\textbf{a}) and $0.540\pm 0.070$ (\textbf{b}) for the case of $t=1.0$, and are $0.095\pm 0.16$ (\textbf{c}) and $0.526\pm 0.014$ (\textbf{d}) for the case of $t=0.5$, respectively.}
	\label{f2}
\end{figure}

In the experiment, we implement a topological photonic lattice consisting of 10 waveguides with $t=1.0$. To perform the evolution-distance average, we integrate 40 such photonic lattices on a single chip with different evolution distances varying from 20 mm to 30 mm with a step size of 0.2mm. We excite one waveguide in the central unit cell $(x=3)$ with a narrowband coherent light at 852 nm, and measure the output density from each photonic lattice. The evolution-distance-dependent PPDC is extracted and shown in Fig.\ref{f2}(a-b). The result in Fig.\ref{f2}(a) shows that, when the system is in the topological trivial phase, the values of PPDC $\bar{P}_d$ keep oscillating centered at $0.045\pm 0.090$. While the system is in the topological nontrivial phase, $\bar{P}_d$ keeps oscillating around $0.540\pm 0.070$ as shown in Fig.\ref{f2}(b). According to Eq. (\ref{wnPPDC}), we obtain the topological winding numbers $\nu=0.09\pm 0.18$ and $\nu=1.08\pm 0.14$ for the two phases respectively. We can see that the oscillation of the measured $P_d$ values is more irregular than that of the simulated result, but the winding number can still be clearly extracted.

\begin{figure}[!t]
	\centering
	\includegraphics[width=0.95\columnwidth]{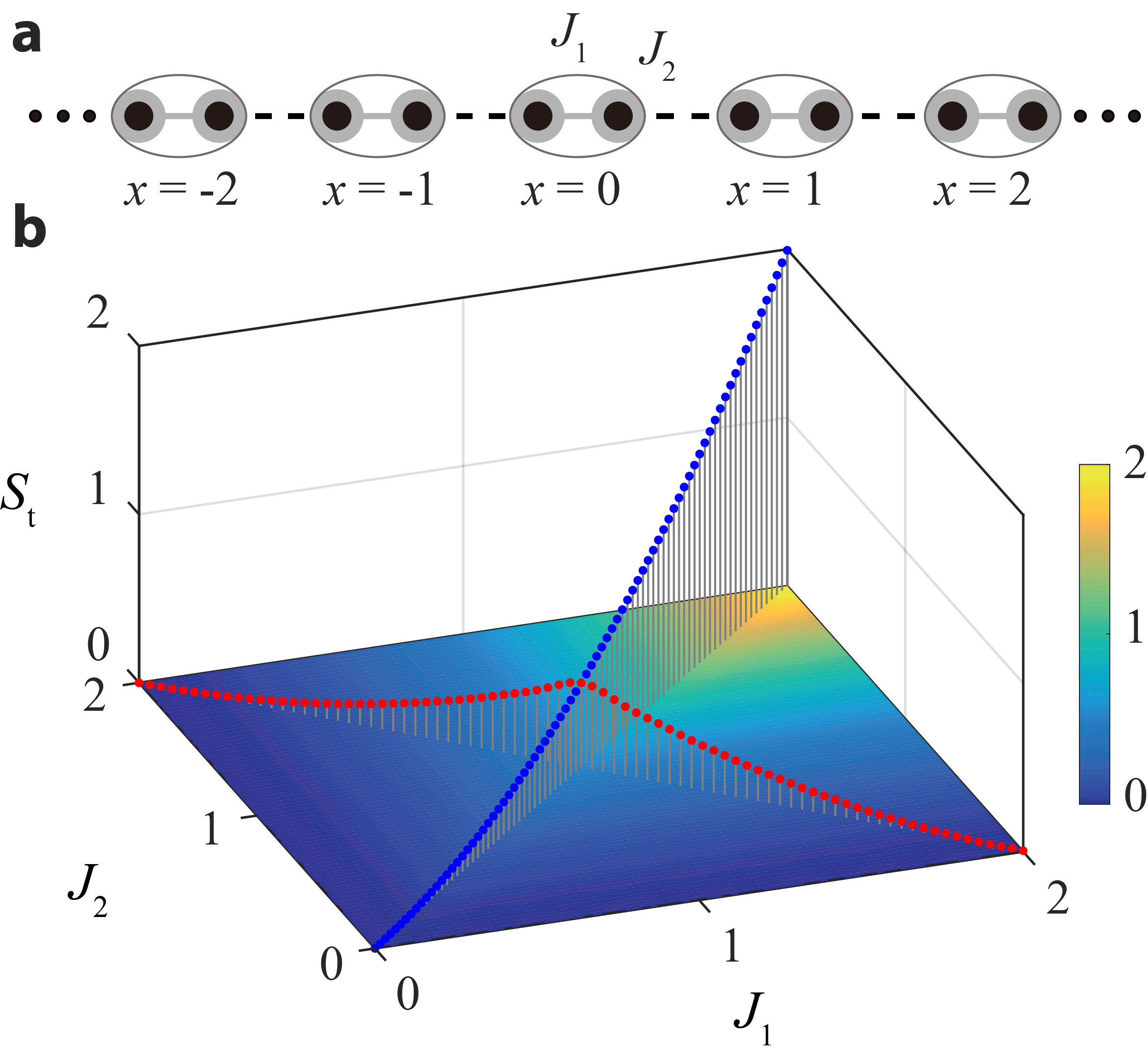}\\
	\caption{\textbf{Theoretical results of TPTS.} \textbf{a,} The Su-Schrieffer-Heeger model with the labels marked starting from middle of the system for a concise expression of the photon population center. \textbf{b,} The simulated results of TPTS. The dynamical TPTS value increases and then decreases for a continuous transitive system (red point), and the transition point will be more distinct in the strong interaction region (blue point).}
	\label{f3}
\end{figure}

\begin{figure}[!t]
	\centering
	\includegraphics[width=0.98\columnwidth]{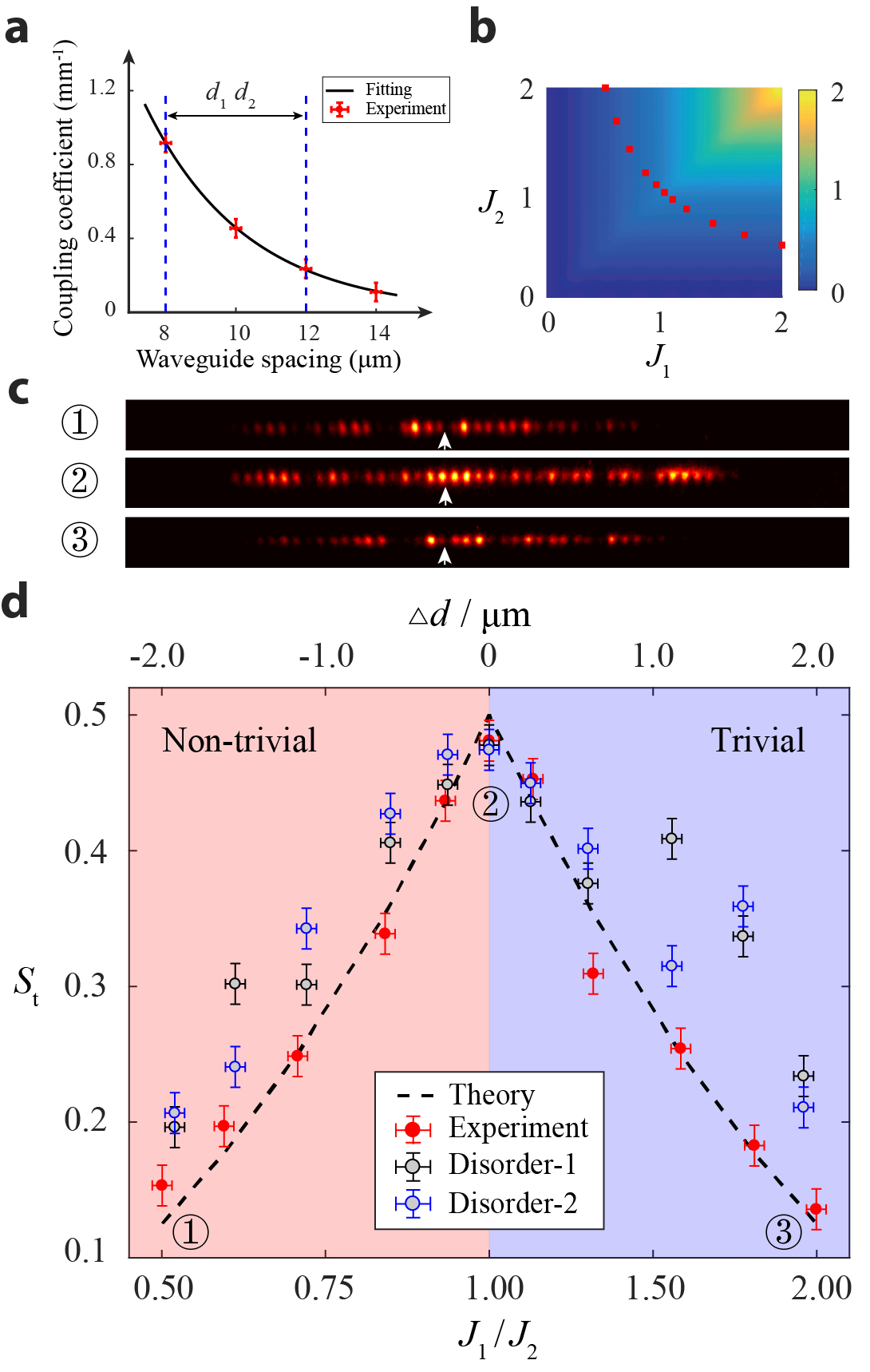}
	\caption{\textbf{Experimental results of TPTS.} \textbf{a,} The relation between the coupling strength and the separation between adjacent waveguides. The blue dash lines mark the experimentally accessible range of $d_1$ and $d_2$. \textbf{b,} The coupling strength of $J_1$ and $J_2$ of the 11 lattices used in experiment (red squares). \textbf{c,} The evolution probability distribution of single photons in topological nontrivial phase, transition point and trivial phase. The blank arrows mark the excited sites in experiment. \textbf{d,} The measured results of TPTS. The topological transition point appears when the system undergoes the phase transition from the topological nontrivial to trivial phase (red circles), even with artificially introduced disorder (black and blue circles).}
	\label{f4}
\end{figure}

To further experimentally demonstrate the reliability and universality of our approach, we fabricate another set of photonic lattices consisting of 18 waveguides with $t=0.5$. The evolution distance varies from 7 mm to 16 mm with a step size of 0.2 mm. As is shown in Fig.\ref{f2}(c-d), when the lattices are prepared in the topological trivial and nontrivial phases, the measured values of $\bar{P_d}$ are oscillating around $0.095\pm 0.16$ and $0.526\pm 0.014$, which lead to the topological winding numbers of $\nu=0.19\pm 0.32$ and $\nu=1.052\pm 0.28$, respectively. The results are well consist with the simulated results shown in Fig.\ref{f1}(c) and suggest that our proposed dynamical approach of measuring topological invariants is insensitive to the detailed lattice configurations.\\


\textbf{Dynamical detection of topological phase transition}.
We further extend the topological system and measurement into quantum regime by observing the topological phase transition point via single-photon dynamics. The topological phase transition in our photonic lattices can also be directly measured from the output density distribution. The transition point can be revealed by the generalized photon population center $P_c=\sum_{x}x^2(a_x^+a_x+b_x^+b_x)$, where the label $x$ is marked as shown in Fig.\ref{f3}(a) for a concise expression. With the bulk state excited from the central unit cell by single photons, the value of generalized photon population center can be derived as $\bar{P}_c(z)$ for an evolution distance $z$ (see Methods). We can further obtain the topological phase transition signal (TPTS) $S_t=\bar{P}_c(z)/z^2$, and we find that (see Methods)
\begin{equation}\label{eq3}
	S_t=
	\begin{cases}
		\frac{J^2_{1}}{2}, & \mbox{$J_{1}<J_{2}$} \\
		\frac{J^2_{2}}{2}. & \mbox{$J_{1}>J_{2}$}
	\end{cases}
\end{equation}

The simulated results are illustrated in Fig.\ref{f3}(b). For a continuously transitive system from topological non-trivial to trivial, as is sketched with the red points, the TPTS value increases firstly and then decreases, and the maximum value arises when the system undergoes the topological phase transition point. Unlike the statistical measurement of PPDC, this approach requires only single measurement on generalized photon population center for a certain structure, the topological transition point therefore can be more conveniently observed in experiment. When the dynamical TPTS value varies with the increasing of the value of $J_1$ and $J_2$, the transition points as sketched with blue points will be more distinct to be observed in the strong interaction region.

To experimentally observe the TPTS in a continuous transitive system, we fabricate 11 set of photonic lattices with lattice constant d=20 $\mu$m and different intra-cell ($d_1$) and inter-cell ($d_2$) space. The dimerization $\varDelta d=(d_2-d_1)/2$ varies from -2 $\mu$m to 2 $\mu$m in a step of 0.5 $\mu$m for 9 lattices, and two more lattices are designed near the transition point with the dimerization values of -0.2 $\mu$m and 0.2 $\mu$m. The coupling strength is modulated by the separation between adjacent waveguides, which is not linear according to coupling mode theory (see Fig.\ref{f4}(a)). The corresponding coupling strength of $J_1$ and $J_2$ defined by the dimerization values in 11 lattices are marked with red squares in Fig.\ref{f4}(b). All the lattices consist of 42 sites (21 untie cells) and have an evolution distance of 18 mm to ensure that the photons will not evolve to the edge. We prepare heralded single photons at 810 nm via spontaneous parametric down conversion (see Methods) and excite the bulk state from one of two sites in the central unit cell $(x=0)$.

We show the experimental results in Fig.\ref{f4}(c-d). Directly from the evolution probability distribution of single photons, we can not find distinct criteria to distinguish when the lattice is in topological nontrivial phase, transition point and trivial phase (see Fig.\ref{f4}(c)). In Fig.\ref{f4}(d), we plot the experimental (red dots) and theoretical (black dash line) results of TPTS $S_t$ varying with the coupling strength ratio $J_1/J_2$ (and dimerization $\varDelta d$). The values of $S_t$ increase with the square of $J_1$ when $J_1/J_2<1$, and decrease with the square of $J_2$ when  $J_1/J_2>1$. As a result, the signal of the topological transition point appears very clearly when the $J_1/J_2=1$, corresponding to the $\varDelta d=0$.

Besides the advantage of directly observing topological transition point, it would be also interesting to test the robustness of this approach. We manage to introduce the disorder into the system by adding random fluctuation of $\pm 0.1\ \mu$m to $d_1$ and $d_2$ in the laser writing process. We fabricate 11 set of such disorder-embedded lattices and repeat the experiment twice. As is shown in Fig.\ref{f4}(c), the experimental results retrieved from the 22 lattices indicate that while the measured values of $S_t$ randomly deviate from the theoretical curve assumed for the ideal case, the topological transition point still can be clearly identified around $J_1/J_2=1$.

In summary, we experimentally demonstrate a direct observation of the topological invariants and the phase transition from the photon dynamics in the bulk state. Our approach provides a new route to direct measurement of topology via single-particle dynamics in the real space, which complements the approach in ultracold atomic systems using Bloch state dynamics in the momentum space~\cite{ZPColdAtoms,BCColdAtoms}. Our approach is also available for further generalization and application to other topological systems and higher-dimensional cases.

The demonstrated key elements, including integrated topological structures, direct measurement in single-photon regime and strong robustness against disorder, can enrich the emerging field of `quantum topological photonics'. With the primary attempt to combine topology with quantum integrated photonics, it is promising to explore scalable topologically protected quantum information processing on topological photonic chips beyond classical topological photonics. The prompt questions, but remain open, will be whether we can directly observe topology with multi-photon dynamics and how qubit and entanglement behave.

\subsection*{Acknowledgments}
The authors thank Jian-Wei Pan for helpful discussions. This research is supported by the National Key Research and Development Program of China (2016YFA0301803, 2017YFA0303700, 2017YFA0304203), National Natural Science Foundation of China (NSFC) (Grant No. 61734005, 11761141014, 11690033, 11604392), Science and Technology Commission of Shanghai Municipality (STCSM) (15QA1402200, 16JC1400405, 17JC1400403), Shanghai Municipal Education
Commission (SMEC)(16SG09, 2017-01-07-00-02-E00049), PCSIRT (IRT 17R70) and the Shanxi 1331KSC and 111 Project (D18001), X.-M.J. acknowledges support from the National Young 1000 Talents Plan.\\

\subsection*{Methods}
{\bf Fabrication and measurement of the integrated topological photonic lattices:} According to the characterized coupling coefficients modulated by the separation between two adjacent waveguides, the lattices are designed and written in borosilicate glass (refractive index $n_0=1.514$) with femtosecond laser with repetition rate 1MHz, pulse duration 290fs and working wavelength 513nm. Before the laser writing beam is focused inside the borosilicate substrate with a 50X objective lens (numerical aperture of 0.55), we control the shape and size of the focal volume of the beam with a beam-shaping cylindrical lens. A high-precision three-axis motion stage is used to move the photonic chip during fabrication with a constant velocity of 5 mm/s.

Experiments are performed by injecting the photons (herald single photon) into the lattices using a 20X objective lens. The evolution output is observed using a 10X microscope objective lens and the CCD (ICCD) camera after a total evolution distance within the lattice.

~\\

\textbf{The relationship between topological winding number and PPDC:}
Suppose one waveguide in the middle unit cell is initially excited, then the initial state of the photonic chip can be written as $|\psi(0)\rangle$. To find the relationship
between the photon dynamics in the waveguide lattice and the winding number, we
introduce population difference in each unit cell and define a
photon population difference center (PPDC), i.e.,
\begin{equation}
	P_{d}=\sum_{x=1}^{N}x(P_{a_{x}}^{e}-P_{b_{x}}^{e}),
\end{equation}%
where $P_{m}^{e}=|e\rangle _{m}\langle e|$ ($m=a_{x},b_{x}$) is the photon population probability. The PPDC associated with the photonic evolution in the waveguide lattice can be described as
\begin{equation}
	\bar{P}_{d}(z)=\langle \psi(0)|e^{iHz}P_{d}e^{-iHz}|\psi(0)\rangle .
\end{equation}%
For the photonic SSH model, we can connect the
above dynamical center to the winding number. We rewrite the PPDC in the momentum space as
\begin{equation}
	\bar{P}_{d}(z)=\frac{1}{2\pi }\int_{-\pi }^{\pi }dk_{x}\langle \chi
	(0)|e^{ih(k_{x})z}i\partial _{k_{x}}\tau _{z}e^{-ih(k_{x})z}|\chi
	(0)\rangle .  \label{DC}
\end{equation}%
By substituting $h(k_x)=d_x\sigma_x+d_y\sigma_y$ into (\ref{DC}), we find that $\bar{P}_{d}(Z)$ can be connected with the topological winding number $\nu$ defined, i.e.,
\begin{equation}
	\begin{split}
		\bar{P}_{d}(z)&=\frac{\nu }{2}-\frac{1}{4\pi}\int dk_{x}\cos(2Ez)\mathbf{n}\times \partial _{k_{x}}\mathbf{n},
		\label{pdk}
	\end{split}
\end{equation}
where $\mathbf{n}=(n_{x},n_{y})=(d_{x},d_{y})/E$ and $E=\sqrt{J_{1}^2+J_{2}^2+2J_{1}J_{2}\cos(k_x)}$. In the long evolution distance limit, the second term in the above equation will vanish. Then we can obtain a relationship between the winding number and the evolution-distance-averaged PPDC, i.e.,
\begin{equation}
	\nu=2\,\mathbb{P}_d,
	\label{pdwn}
\end{equation}%
where the evolution-distance-averaged PPDC is
\begin{equation}
	\mathbb{P}_d={\lim_{Z\rightarrow \infty }}\frac{1}{Z}\int_{0}^{Z}dz\,\bar{P}_{d}(z).
\end{equation}
where $Z$ is the total evolution distance for the photons in the waveguide lattice. Note that $\mathbb{P}_d$ is just the oscillation center of $\bar{P}_d(z)$ varying with $z$. The topological winding number is twice this oscillation center.

~\\

\textbf{The relationship between TPTS and photon population center:}
The initial state of the photonic chip is the same as the one in the winding number detection. To find the relation between the photon dynamics in the waveguide lattice and the winding number, we
introduce a generalize photon population center operator, i.e.,
\begin{equation}
	P_{c}=\sum_{x=1}^{N}x^2(P_{a_{x}}^{e}+P_{b_{x}}^{e}),
\end{equation}%
where $P_{m}^{e}=|e\rangle _{m}\langle e|$ ($m=a_{x},b_{x}$) is the photon population operator. Then the generalized photon population center associated with the evolution of photons in the waveguide lattice can be described as
\begin{equation}
	\bar{P}_{c}(z)=\langle \psi(0)|e^{iHz}P_{c}e^{-iHz}|\psi(0)\rangle .
\end{equation}%
Based on the above equation, we define a new quantity called as topological phase transition signal (TPTS), which is expressed as
\begin{equation}
	S_t=\bar{P}_{c}(z)/z^2.
\end{equation}
By transferring the above equation into the momentum space, we can further get
\begin{equation}
	S_t=\frac{1}{2\pi }\int_{-\pi }^{\pi }dk_{x}\langle \psi (0)|e^{i%
		\hat{h}(k_{x})z}(i\partial _{k_{x}})^2e^{-i\hat{h}(k_{x})z}|\psi (0)\rangle/z^2.
\end{equation}
In the long evolution distance limit, the terms proportional to $1/z$ can be omitted and the above identity can be simplified into
\begin{eqnarray}
	S_t&=&\frac{1}{2\pi}\int_{-\pi }^{\pi }dk_{x}(\partial _{k_{x}}E)^2 \nonumber \\
	&=&\frac{1}{2\pi }\int_{-\pi }^{\pi }dk_{x}\frac{J^2_{1}J^2_{2}\sin^2(k_x)}{J^2_{1}+J^2_{2}+2J^2_{1}J^2_{2}\cos(k_x)}.
	\label{yt}
\end{eqnarray}
Based on residue theorem, we can analytically solve the above integral and get
\begin{equation}
	S_t=%
	\begin{cases}
		\frac{J^2_{1}}{2}, & \mbox{$J_{1}<J_{2}$} \\
		\frac{J^2_{2}}{2}. & \mbox{$J_{1}>J_{2}$}%
	\end{cases}%
\end{equation}
This equation shows that the topological phase transition in the photonic Su-Schrieffer-Heeger model can be directly observed from the single photon dynamics in bulk state.

~\\

{\bf The generation and imaging of the heralded single photons:} We obtain the single-photon source with the wavelength of 810 nm generated from periodically-poled KTP (PPKTP) crystal via spontaneous parametric down conversion (SPDC). After a long-pass filter and a polarized beam splitter (PBS), the photon pairs are separated to two components, horizontal and vertical polarization. The measured evolution patterns would come from the thermal-state light rather than single-photons if we inject only one polarized photon into the lattices without external trigger. Therefore, we inject the horizontally polarized photon into the lattices, while the vertically polarized photon acts the trigger for heralding the horizontally polarized photons out from the lattices with a time slot of 10ns. We capture each evolution result using ICCD camera after accumulating in the `external' triggering mode for 2000s.

\end{document}